\documentclass[10pt]{article}
\usepackage{oldgerm}
\usepackage{euler}
\usepackage{graphics}
\usepackage{bm}
\usepackage{graphicx}
\usepackage{amsmath}
\usepackage{amssymb}
\usepackage{amstext}
\usepackage{amscd}
\usepackage{amsfonts}
\usepackage{appendix}
\usepackage{color}
\DeclareMathAlphabet{\EuFrak}{U}{euf}{m}{n}
\DeclareMathAlphabet{\EuScript}{U}{eus}{m}{n}

\newcommand{\nd}{\noindent}

\newcommand{\be}{\begin{equation}}
\newcommand{\ee}{\end{equation}}
\newcommand{\ben}{\begin{eqnarray}}
\newcommand{\een}{\end{eqnarray}}

\title{{\bf Physical peculiarities of divergences emerging
in q-deformed statistics}}

\author{{M. C. Rocca$^1$, A. Plastino$^1$,  and  G. L. Ferri $^2$} \\
\small{$^1$ La Plata National University and
  Argentina's National Research Council}\\
\small{(IFLP-CCT-CONICET)-C. C. 727, 1900 La Plata - Argentina}\\
$^2$ \small{Fac. de C. Exactas-National University La Pampa,} \\
\small{Peru y Uruguay, Santa Rosa, La Pampa, Argentina}}

\begin{document}

\maketitle

\begin{abstract}

\nd It was found in [Europhysics Letters {\bf 104}, (2013), 60003]
that classical Tsallis theory exhibits poles in the partition
function  ${\cal Z}$ and the mean energy $<{\cal U}>$. These occur
 at a countably set of the q-line. We give here, via a simple procedure, a mathematical account of them.
 Further, by focusing
attention upon the pole-physics, we encounter interesting effects.
In particular, for the specific heat, we uncover hidden
gravitational effects.

\nd K-WORDS: Tsallis entropy, divergences,  partition function,
specific heat.

\end{abstract}

\newpage

\renewcommand{\theequation}{\arabic{section}.\arabic{equation}}

\section{Introduction}

\setcounter{equation}{0}

\nd Generalized or q-statistical mechanics \`a la Tsallis has
generated manifold applications in the last 25 years
\cite{tsallis,web,epjb1,epjb2,epjb3,epjb4,epjb5,epjb6,
epjb7,epjb8,epjb9}. It has been shown (see for instance,
\cite{PP93,chava}) that the Tsallis q-statistics is of great
importance for dealing with some astrophysical issues involving
self-gravitating systems \cite{lb}. Moreover, this statistics has
proved its utility in variegated scientific fields, with several
thousands publications and authors \cite{web}, so that studying
its structural features is an important issue for physics,
astronomy, biology, neurology, economics, etc. \cite{tsallis}. The
success of the q-statistics reaffirms the well grounded notion
asserting that there is much physics whose origin is of purely
statistical nature (not mechanical). As a spectacular example, me
mention the application of q-ideas to high energy experimental
physics, where the q-statistics appears to adequately describe the
transverse momentum distributions of different hadrons
\cite{tp11o,tp11,phenix}. 

\nd In this work we show that as yet unexplored  gravitational
effects characterize this q-theory on account of {\it divergences}
that, in some circumstances, emerge, within the q-statistical
framework,  in both the mean energy and the partition function.

\nd Divergences are an important topic in theoretical physics.
Indeed, the study and elimination of divergences of a physical
theory is perhaps one of the most important aspects of theoretical
physics. The quintessential typical example is the attempt to
quantify the gravitational field, which so far has not been
achieved. Some examples of elimination of divergences can be seen
in  references \cite{tq1,tq2,tq3,tq4,tq5}.

\nd  We will use here  an extremely  simplified version of the
ideas of \cite{tq1,tq2,tq3,tq4,tq5} in connection with  Tsallis
q-statistics \cite{tsallis,web}, with emphasis in its
applicability to gravitational  issues \cite{PP93,chava}, in
particular  self-gravitating systems \cite{lb}.
 We will see that the removal of the above mentioned divergences
 leads to illuminating insights.

\section{The divergences of q-statistics}

\setcounter{equation}{0}

As we have shown in \cite{epl}, the q-partition function of the
{\it classical} Harmonic Oscillator (HO) in $\nu$ dimensions can
be written in the form
\begin{equation}
\label{ep2.1} {\cal Z}=\frac {\pi^{\nu}} {\Gamma(\nu)}
\int\limits_0^{\infty} \frac {u^{\nu-1}} {[1+\beta(q-1)u]^{\frac
{1} {q-1}}} du,
\end{equation}
where $u$ refers to the phase space energy and $\beta$ is the
inverse temperature. The result of integral (\ref{ep2.1}) is,
according to \cite{grad},
\begin{equation}
\label{ep2.2}
{\cal Z}=
\frac {\pi^{\nu}} {[\beta(q-1)]^{\nu}}
\frac {\Gamma\left(\frac {1} {q-1}-\nu\right)}
{\Gamma\left(\frac {1} {q-1}\right)}
\end{equation}
This result is valid for $q\neq 1$ and we have selected $1\leq q
<2$. Of course,  $q=1$ is the  Boltzmann statistics instance, for
which the q-exponential transforms itself into the ordinary
exponential function (and the integral (\ref{ep2.1}) is
convergent). According to (\ref{ep2.2}), the singularities
(divergences) of (\ref{ep2.1}) are given by the poles of the
$\Gamma$ function that appears in the numerator of (\ref{ep2.2}),
i.e., for
\[\frac {1} {q-1}-\nu=-p\;\;{\rm for} \;\;p=0,1,2,3,......,\]
or, equivalently, for
\[q=\frac {3} {2},\frac {4} {3},\frac {5} {4},\frac {6} {5},......,
\frac {\nu} {\nu-1},\frac {\nu+1} {\nu}\] In a similar way, we
have for the q-mean energy of the HO,
\begin{equation}
\label{ep2.3}
<{\cal U}>=\frac {\pi^{\nu}} {\Gamma(\nu){\cal Z}}
\int\limits_0^{\infty} \frac {u^{\nu}}
{[1+\beta(q-1)u]^{\frac {1} {q-1}}} du
\end{equation}
The result of (\ref{ep2.3}) is,  using \cite{grad} once again,
\begin{equation}
\label{ep2.4} <{\cal U}>=\frac {\nu\pi^{\nu}} {{\cal
Z}[\beta(q-1)]^{\nu+1}} \frac {\Gamma\left(\frac {1}
{q-1}-\nu-1\right)} {\Gamma\left(\frac {1} {q-1}\right)},
\end{equation}
where we assume that ${\cal Z}$ is the physical partition
function, which has no singularities. In this case, the
singularities of (\ref{ep2.4}) are given by:
\[\frac {1} {q-1}-\nu-1=-p\;\;\;{\rm for}\;p=0,1,2,3,......,\]
or, equivalently,
\[q=\frac {3} {2},\frac {4} {3},\frac {5} {4},\frac {6} {5},......,
\frac {\nu+1} {\nu},\frac {\nu+2} {\nu+1}.\] As usual
\cite{2PP93}, in terms of the so-called q-logarithms
\cite{tsallis} $\ln_q(x)= \frac{x^{1-q} - 1 }{1-q}$, the entropy
is cast in the fashion
\begin{equation}
\label{ep2.5}
S=\ln_q {\cal Z} + {\cal Z}^{1-q} \beta <{\cal U}>
\end{equation}
and it is finite if ${\cal Z}$ and $<{\cal U}>$ are also finite.

\nd Our purpose here is then to derive, for the classical HO,
physical thermo-statistical variables ${\cal Z}$, $<{\cal U}>$,
and ${\cal S}$, by appropriately treating (regularizing) the above
singularities. As an illustration, we specify things for the cases
of dimensions one, two,  three, and $N$.

\section{The one-dimensional case}

\setcounter{equation}{0}

In one dimension ${\cal Z}$ is regular and $<{\cal U}>$ has a
singularity at $q=\frac {3} {2}$. For $q\neq\frac {3} {2}$, ${\cal
Z}$ and $<{\cal U}>$ can be easily evaluated. The result is
straightforward
\begin{equation}
\label{ep3.1} {\cal Z}= \frac {\pi} {\beta(2-q)},
\end{equation}
\begin{equation}
\label{ep3.2} <{\cal U}>=\frac {1} {\beta(3-2q)}.
\end{equation}
According to (\ref{ep3.2}), in the regular case, as $<{\cal
U}>\geq 0$, one should have $q<\frac {3} {2}$. At $q=3/2$ we have
a pole in the mean energy, that we wish to investigate. Instead,
when $q=\frac {3} {2}$, we have for ${\cal Z}$
\begin{equation}
\label{ep3.4} {\cal Z}= \frac {2\pi} {\beta},
\end{equation}
a regular value.  Regularization is needed then only for $<{\cal
U}>$.

\subsection{Dealing with the divergence}

In order to proceed with such regularizing procedure, the main
idea is to  write $<{\cal U}>$ as a function of the dimension
$\nu$, in the fashion
\begin{equation}
\label{ep3.5} <{\cal U}>=\frac {2^{\nu+1}\nu\pi^{\nu}} {{\cal
Z}{\beta}^{\nu+1}} \Gamma(1-\nu),
\end{equation}
and carefully dissect this expression.  First we  recast things as

\begin{equation}
\label{ep3.5bis} <{\cal U}>=\frac {2^{\nu+1}[\nu-1+1]\pi^{\nu}}
{{\cal Z}{\beta}^{\nu+1}} \Gamma(1-\nu),
\end{equation}
and remember that $(\nu-1) \Gamma(1-\nu)= -\Gamma(2-\nu)$ to
obtain

\begin{equation}
\label{ep3.6} <{\cal U}>=-\frac {1} {\pi{\cal Z}} \left(\frac
{2\pi} {\beta}\right)^{\nu+1} \Gamma(2-\nu)+ \frac {1} {\pi{\cal
Z}} \left(\frac {2\pi} {\beta}\right)^2 \left(\frac {2\pi}
{\beta}\right)^{\nu-1}\Gamma(1-\nu).
\end{equation}
We realize that the first term of (\ref{ep3.6}) is finite, while
the second one is singular for $\nu=1$ (the physical dimension in
this instance is unity). The trick here is to appeal to a Taylor's
expansion, around $\nu=1$, {\it of the third factor in the second
term}, i.e., $\frac {2\pi} {\beta}^{\nu-1}=
\exp{[(\nu-1)\ln{\frac{2\pi}{\beta}}]}$. Notice also that, from
(\ref{ep3.4}), ${\cal Z}= \frac {2\pi} {\beta}$. Accordingly, we
have
\[<{\cal U}>=-\frac {1} {(2\pi^2/\beta)}
\left(\frac {2\pi} {\beta}\right)^{\nu+1} \Gamma(2-\nu)+
(2/\beta)\times\]
\begin{equation}
\label{ep3.7} \left[1+(\nu-1)\ln\left(\frac {2\pi} {\beta}\right)+
\frac {(\nu-1)^2} {2}{\ln}^2\left(\frac {2\pi} {\beta}\right)+
\cdot\cdot\cdot\right]\Gamma(1-\nu).
\end{equation}
We use now once again the fact that  $(\nu-1)\Gamma(1-\nu) =
-\Gamma(2-\nu)$ to write

\[<{\cal U}>=-\frac {1} {(2\pi^2/\beta)}
\left(\frac {2\pi} {\beta}\right)^{\nu+1} \Gamma(2-\nu)+
(2/\beta)\times\]
\begin{equation}
\label{ep3.7bis} \left[1-\ln\left(\frac {2\pi} {\beta}\right)-
\frac {(\nu-1)} {2}{\ln}^2\left(\frac {2\pi} {\beta}\right)+
\cdot\cdot\cdot\right]\Gamma(2-\nu),
\end{equation}
and then, in the limit $\nu \rightarrow 1$, after cancellations
and series' terms  that vanish, we are left with

\begin{equation}
\label{ep3.8} <{\cal U}>=-\frac {2} {\beta} \left[1+\ln\left(\frac
{2\pi} {\beta}\right)\right],
\end{equation}
that is to be regarded as the physical value of $ <{\cal U}>$
\cite{tq1,tq2,tq3,tq4,tq5,xxx}.

\nd Using now (\ref{ep2.5}) we immediately get for ${\cal S}$
\begin{equation}
\label{ep3.9}
{\cal S}=\ln_{\frac {3} {2}}\left(\frac {2\pi} {\beta}\right)
-\sqrt{\frac {2\pi} {\beta}}
\left[1+\ln\left(\frac {2\pi} {\beta}\right)\right]
\end{equation}

\subsection{Direct proof of the existence of an upper bound to the canonical bath' temperature}
\nd Since the mean energy must be positive,  according to
(\ref{ep3.8}) the possible values of $\beta$ are restricted by the
constraint $\beta>2\pi e$, entailing $T< 1/2\pi ek_B$, with $k_B$
Boltzmann's constant. There is an upper bound to the physical
temperature, which cannot be infinite. This agrees with the
considerations made in \cite{PP94}: {\it q-statistics refers to
systems in thermal contact with a {\bf finite} bath}.

\subsection{A fancier conjecture}
\nd On a more conjectural fashion, one is also reminded here of
the Hagedorn temperature. This is the temperature at which
ordinary matter is no longer stable and would evaporate,
transforming itself into quark matter, a sort of boiling point of
hadronic matter. This temperature would exist on account of the
fact that the accessible energy would be so high that
quark-antiquark pairs would be be spontaneously extracted from the
vacuum. A putative system at such a high temperature is able to
accommodate any  amount of energy because the newly emerging
quarks would provide additional degrees of freedom. The Hagedorn
 temperature would thus be unsurmountable \cite{witten}.

\section{The two-dimensional case}

\setcounter{equation}{0}

\nd For two dimensions, ${\cal Z}$ has a singularity at $q=\frac
{3} {2}$ and $<{\cal U}>$ has singularities at $q=\frac {3} {2}$
and $q=\frac {4} {3}$. Save for the case of   these singularities,
we can evaluate their values of the main statistical quantities
without the use of dimensional regularization. Thus, we obtain
\begin{equation}
\label{ep4.1} {\cal Z}=\frac {\pi^2} {\beta^2(2-q)(3-2q)},
\end{equation}
\begin{equation}
\label{ep4.2} <{\cal U}>=\frac {2} {\beta(4-3q)},
\end{equation}
\begin{equation}
\label{ep4.3}
{\cal S}=\ln_q\left[\frac {\pi^2}
{\beta^2(2-q)(3-2q)}\right]+
\left[\frac {\pi^2}
{\beta^2(2-q)(3-2q)}\right]^{1-q}
\frac {2} {4-3q}
\end{equation}
According to (\ref{ep4.2}), in the regular  case $q<\frac {4}
{3}$.

\subsection{The $q=3/2$ pole}

\noindent For $q=\frac {3} {2}$ we must employ the treatment of
the preceding Section, i.e.,  regularize, both  ${\cal Z}$ and
${\cal U}$. We start with ${\cal Z}$. From (\ref{ep2.2}) we have
\begin{equation}
\label{ep4.4} {\cal Z}=\left(\frac {2\pi}
{\beta}\right)^{\nu}\Gamma(2-\nu),
\end{equation}
which can be rewritten as
\begin{equation}
\label{ep4.5} {\cal Z}=\left(\frac {2\pi} {\beta}\right)^2
\left(\frac {2\pi} {\beta}\right)^{\nu-2}\Gamma(2-\nu).
\end{equation}
With this form for ${\cal Z}$, we can expand  in Taylor's series,
around $\nu=2$, the factor $\left(\frac {2\pi}
{\beta}\right)^{\nu-2}= \exp{[(\nu-2)\ln{\frac {2\pi} {\beta}}]}$,
noting also that $(\nu-2)\Gamma(2-\nu)=-\Gamma(3-\nu)$, i.e.,
\begin{equation}
\label{ep4.6} {\cal Z}=\left(\frac {2\pi} {\beta}\right)^2
\Gamma(2-\nu)\left[1+(\nu-2)\ln\left(\frac {2\pi} {\beta}\right)
\cdot\cdot\cdot\right],
\end{equation}
and thus we obtain the physical value of ${\cal Z}$ as
\begin{equation}
\label{ep4.7} {\cal Z}=-\frac {4\pi^2} {\beta^2} \ln\left(\frac
{2\pi} {\beta}\right).
\end{equation}

\noindent For ${\cal U}$ the situation is similar. From
(\ref{ep2.4}) we have
\begin{equation}
\label{ep4.8} <{\cal U}>=\frac {\nu} {{\cal Z}\pi} \left(\frac
{2\pi} {\beta}\right)^{\nu+1}\Gamma(1-\nu),
\end{equation}
where ${\cal Z}$ is given by (\ref{ep4.7}). Proceeding in the same
way as we did in the one dimensional case, and omitting here from
intermediate steps,  we rewrite $<{\cal U}>$ in the fashion
\begin{equation}
\label{ep4.9} <{\cal U}>=\frac {\Gamma(3-\nu)} {{\cal
Z}\pi(\nu-1)} \left(\frac {2\pi} {\beta}\right)^{\nu+1}+ \frac {2}
{{\cal Z}\pi} \left(\frac {2\pi} {\beta}\right)^3 \left(\frac
{2\pi} {\beta}\right)^{\nu-2} \frac {\Gamma(2-\nu)} {1-\nu},
\end{equation}
and we obtain the physical value of $<{\cal U}>$:
\begin{equation}
\label{ep4.10} <{\cal U}>=\frac {8\pi^2} {{\cal Z}\beta^3}+ \frac
{16\pi^2} {{\cal Z}\beta^3}\ln\left(\frac {2\pi} {\beta}\right),
\end{equation}
so that  replacing $Z$ by the value given in (\ref{ep4.7}) we have
\begin{equation}
\label{ep4.11} <{\cal U}>=\frac {2} {\beta(\ln\beta-\ln 2\pi)}+
\frac {4} {\beta}.
\end{equation}
From (\ref{ep4.11}) we see that the possible values of $\beta$ are
 given by $\beta>2\pi$. {\it Again, a temperature's upper bound is detected.}

\nd  Now, from the physical values of ${\cal Z}$ and $<{\cal U}>$,
as given by (\ref{ep4.7}) and (\ref{ep4.11}), respectively, and
from (\ref{ep2.5}), we find the physical value of ${\cal S}$ as
\begin{equation}
\label{ep4.12}
{\cal S}=\ln_{\frac {3}
{2}}\left[\frac {4\pi^2} {\beta^2} \ln\left(\frac {\beta}
{2\pi}\right)\right]+\left[\frac {4\pi^2} {\beta^2}
\ln\left(\frac {\beta} {2\pi}\right)\right]^{
-\frac {1} {2}}
\left[\frac {2}
{(\ln\beta-\ln 2\pi)}+4\right]
\end{equation}

\subsection{The $q=4/3$ pole}

\nd For $q=\frac {4} {3}$, ${\cal Z}$ is finite and $<{\cal U}>$
has a pole. The procedure for finding their physical values is
similar to that for the case $q=\frac {3} {2}$. For this reason,
we
 only indicate the results obtained for ${\cal Z}$, $<{\cal U}>$, and
${\cal S}$. One finds
\begin{equation}
\label{ep4.13} {\cal Z}=\frac {9\pi^2} {2\beta^2},
\end{equation}
\begin{equation}
\label{ep4.14} <{\cal U}>=\frac {6} {\beta}\left[\ln\left(\frac
{\beta} {3\pi}\right) -\frac {1} {2}\right],
\end{equation}
\begin{equation}
\label{ep4.15}
{\cal S}=\ln_{\frac {4}
{3}}\left(\frac {9\pi^2} {2\beta^2}\right)+
\left(\frac {9\pi^2}
{2\beta^2}\right)^{-\frac {1} {3}}
\left[6\ln\left(\frac
{\beta} {3\pi}\right) -3\right]
\end{equation}
From (\ref{ep4.14}) we see that the possible values of $\beta$ are
given by the constraint $\beta>3\pi\sqrt{e}$.

\section{The three-dimensional case}

\setcounter{equation}{0}

\nd In three dimensions, ${\cal Z}$ has poles at $q=\frac {3} {2}$
and $q=\frac {4} {3}$ while $<{\cal U}>$ exhibits them at $q=\frac
{3} {2}$, $q=\frac {4} {3}$, and $q=\frac {5} {4}$. Consequently,
after  regularization,  we have
\begin{equation}
\label{ep5.1} {\cal Z}=\frac {\pi^3} {\beta^3(2-q)(3-2q)(4-3q)},
\end{equation}
\begin{equation}
\label{ep5.2} <{\cal U}>=\frac {3} {\beta(5-4q)}.
\end{equation}
From  (\ref{ep5.1}) and (\ref{ep5.2}) we obtain for the entropy
\begin{equation}
\label{ep5.3}
{\cal S}=
\ln_q\left[\frac {\pi^3} {\beta^3(2-q)(3-2q)(4-3q)}\right]+
\left[\frac {\pi^3} {\beta^3(2-q)(3-2q)(4-3q)}\right]^{q-1}
\frac {3} {5-4q}
\end{equation}
In this case $q$ should satisfy the condition $q<\frac {5} {4}$
for the mean energy to be a positive quantity. 

\subsection{The $q=3/2$ pole}

\noindent For $q=\frac {3} {2}$ we have
\begin{equation}
\label{ep5.4} {\cal Z}=\left(\frac {2\pi} {\beta}\right)^{\nu}
\Gamma(2-\nu).
\end{equation}
Proceeding as in the previous cases and making now the Taylor's
expansion around $\nu=3$, ${\cal Z}$ acquires the appearance
\begin{equation}
\label{ep5.5} {\cal Z}=\left(\frac {2\pi} {\beta}\right)^3 \frac
{\Gamma(3-\nu)} {2-\nu}\left[ 1+(\nu-3)\ln\left(\frac {2\pi}
{\beta}\right)+\cdot\cdot\cdot\right].
\end{equation}
From (\ref{ep5.5}) it is easy to obtain the physical value of
${\cal Z}$ as
\begin{equation}
\label{ep5.6} {\cal Z}=\frac {8\pi^3} {\beta^3} \ln\left(\frac
{2\pi} {\beta}\right).
\end{equation}
In a similar vein have for $<{\cal U}>$
\begin{equation}
\label{ep5.7} <{\cal U}>=\frac {1} {\beta(\ln\beta-\ln 2\pi)}-
\frac {3} {\beta},
\end{equation}
and from (\ref{ep5.6}) and (\ref{ep5.7})
\begin{equation}
\label{ep5.8}
{\cal S}=\ln_{\frac {3} {2}}
\left[\frac {8\pi^3} {\beta^3} \ln\left(\frac {2\pi}
{\beta}\right)\right]+
\left[\frac {8\pi^3} {\beta^3}
\ln\left(\frac {2\pi} {\beta}\right)\right]^{-\frac {1} {2}}
\left(\frac {1} {\ln\beta-\ln 2\pi}-3\right)
\end{equation}
with $2\pi<\beta<2\pi e^{\frac {1} {3}}$.  This entails that the
system exhibits positive entropy  only for a small range of very
high temperatures. 

\subsection{The $q=4/3$ and $q=5/4$ poles}

\noindent For $q=\frac {4} {3}$ and $q=\frac {5} {4}$ we give only
the corresponding results, since the calculations are entirely
similar to those for the case $q=\frac {3} {2}$. Thus, for
$q=\frac {4} {3}$ we have
\begin{equation}
\label{ep5.9} {\cal Z}=\frac {27\pi^3} {2\beta^3} \ln\left(\frac
{\beta} {3\pi}\right),
\end{equation}
\begin{equation}
\label{ep5.10} <{\cal U}>=\frac {3} {\beta(\ln\beta-\ln 3\pi)}-
\frac {9} {\beta},
\end{equation}
\begin{equation}
\label{ep5.11}
{\cal S}=\ln_{\frac {4} {3}}\left[\frac {27\pi^3} {2\beta^3}
\ln\left(\frac {\beta} {3\pi}\right)\right]+
\left[\frac {27\pi^3} {2\beta^3}
\ln\left(\frac {\beta} {3\pi}\right)\right]^{-\frac {1} {3}}
\left(\frac {3} {\ln\beta-\ln 3\pi}-9\right)
\end{equation}
with $3\pi<\beta<3\pi e^{\frac {1} {3}}$.  This entails, again,
that the system exhibits positive entropy  only for a small range
of very high temperatures. 

\noindent
For $q=\frac {5} {4}$:
\begin{equation}
\label{ep5.12} {\cal Z}=\frac {32\pi^3} {3\beta^3},
\end{equation}
\begin{equation}
\label{ep5.13} <{\cal U}>=\frac {12} {\beta} \ln\left(\frac
{\beta} {4\pi}\right)- \frac {4} {\beta},
\end{equation}
\begin{equation}
\label{ep5.14}
{\cal S}=\ln_{\frac {5}
{4}}\left(\frac {32\pi^3} {3\beta^3}\right)+
\left(\frac {32\pi^3}
{3\beta^3}\right)^{-\frac {1} {4}}
\left[12\ln\left(\frac
{\beta} {4\pi}\right)-4\right],
\end{equation}
with $\beta>4\pi e^{\frac {1} {3}}$.

\section{The N-Dimensional Case}

\setcounter{equation}{0}

Repeating the calculation made for 2, 3 and 4 dimensions,
with more algebraic work we get for
$ {\cal Z} $ the expression:
\begin{equation}
\label{ep7.1}
{\cal Z}_{\frac {\nu-k+1} {\nu-k}}=\frac {(-1)^{k+1}} {k!\Gamma(\nu-k)}
\left[\frac {(\nu-k)\pi} {\beta}\right]^{\nu}
\ln\left[\frac {(\nu-k)\pi} {\beta}\right]
\end{equation}
Here $k=0,1,2,3........,\nu-2,$where $\nu$ is the
dimension of the space.
And for $<U>$:
\[<U>_{\frac {\nu-k+2} {\nu-k+1}}=\frac {(-1)^{k+1}}
{k!\Gamma(\nu-k)\beta Z}
\left[\frac {(\nu+1-k)\pi} {\beta}\right]^{\nu}+\]
\begin{equation}
\label{ep7.2}
\frac {(-1)^{k+1}\nu} {k!\Gamma(\nu-k)\beta Z}
\left[\frac {(\nu+1-k)\pi} {\beta}\right]^{\nu}
\ln\left[\frac {(\nu+1-k)\pi} {\beta}\right]
\end{equation}
where $k=0,1,2,3........,\nu-1$.

\section{Specific Heats}

\setcounter{equation}{0}

We set $k\equiv k_B$.  For $\nu=1$, in the regular case we have
for the specific heat $C$:
\begin{equation}
\label{ep6.1} {\cal C}=\frac {k} {3-2q},
\end{equation}
with $q<\frac {3} {2}$.

\noindent For $\nu=2$ one has
\begin{equation}
\label{ep6.2} {\cal C}=\frac {2k} {4-3q},
\end{equation}
with $q<\frac {4} {3}$.

\noindent Finally, for $\nu=3$ one ascertains that
\begin{equation}
\label{ep6.3} {\cal C}=\frac {3k} {5-4q},
\end{equation}
with $q<\frac {5} {4}$.

\subsection{Specific heats at the poles}

\noindent  For $\nu=1$; $q=\frac {3} {2}$
\begin{equation}
\label{ep6.4} {\cal C}=-2k(\ln kT+\ln 2\pi +2).
\end{equation}
with $kT<\frac {1} {2\pi e}$.

\noindent
For $\nu=2$; $q=\frac {3} {2}$
\begin{equation}
\label{ep6.5} {\cal C}=\frac {2k} {(\ln kT+\ln 2\pi)^2}- \frac
{2k} {(\ln kT+\ln 2\pi)}+4k,
\end{equation}
with $kT<\frac {1} {2\pi}$.

\noindent For $\nu=2$ and  $q=\frac {4} {3}$ things become:
\begin{equation}
\label{ep6.6} {\cal C}=-6k\left(\ln kT+\ln 3\pi +\frac {3}
{2}\right),
\end{equation}
with $kT<\frac {1} {3\pi\sqrt{e}}$.

\noindent For $\nu=3$; $q=\frac {3} {2}$,
\begin{equation}
\label{ep6.7} {\cal C}=\frac {k} {(\ln kT+\ln 2\pi)^2}- \frac {k}
{(\ln kT+\ln 2\pi)}-3k,
\end{equation}
with $\frac {1} {2\pi e^{\frac {1} {3}}}<kT<\frac {1} {2\pi}$.

\noindent For $\nu=3$ and $q=\frac {4} {3}$ one has
\begin{equation}
\label{ep6.8} {\cal C}=\frac {3k} {(\ln kT+\ln 3\pi)^2}- \frac
{3k} {(\ln kT+\ln 3\pi)}-9k,
\end{equation}
with $\frac {1} {3\pi e^{\frac {1} {3}}}<kT<\frac {1} {3\pi}$

\noindent Finally, for $\nu=3$ and $q=\frac {5} {4}$
\begin{equation}
\label{ep6.9} {\cal C}=-12k\left(\ln kT+\ln 4\pi +\frac {4}
{3}\right)
\end{equation}
with $kT<\frac {1} {4\pi e^{\frac {1} {3}}}$.

\nd   Figs, 1, 2, and 3 plot the pole-specific heats within their
allowed temperature ranges, for one, two, and three dimensions,
respectively. The most distinguished feature emerges in the cases
in which we deal with $<U>-$poles for which $Z$ is regular. We see
in such a case that negative specific heats arise. Such an
occurrence has been associated to self-gravitational systems
\cite{lb,binney}. In turn, Verlinde has associated this type of
systems to an entropic force \cite{verlinde}. It is natural to
conjecture then that such a force may appear at the energy poles.

\nd Notice also that temperature ranges are restricted. There is
an $T-$upper bound that one may wish to link to the Hagedorn
temperature (see above) \cite{witten}. In two and three dimensions
there is also a lower bound, so that the system (at the poles)
would be stable only in a limited $T-$range.

\section{Discussion}

\setcounter{equation}{0}

\nd In this work we have appealed to an an elementary
regularization procedure to study the poles in the partition
function and the mean energy that appear, for specific, discrete
q-values, in Tsallis' statistics. We studied the thermodynamic
behavior at the poles and found interesting peculiarities. The
analysis was made in one, two, three, and $N$ dimensions. Amongst
pole-traits we emphasize:

\begin{itemize}

\item We have proved that there is an upper bound to the
temperature at the poles, confirming the findings of Ref.
\cite{PP94}.

\item In some cases, Tsallis' entropies are positive only for a
restricted temperature-range.

\item Negative specific heats, characteristic trait of
self-gravitating systems, are encountered.

\end{itemize}

\nd \color{red} Our physical results derive only from statistics,
not from mechanical effects. This fact reminds us of a similar
occurrence in the case of the entropic force conjectured by
Verlinde \cite{verlinde}. \normalcolor

\newpage
\begin{figure}[h]
\begin{center}
\includegraphics[scale=0.6,angle=0]{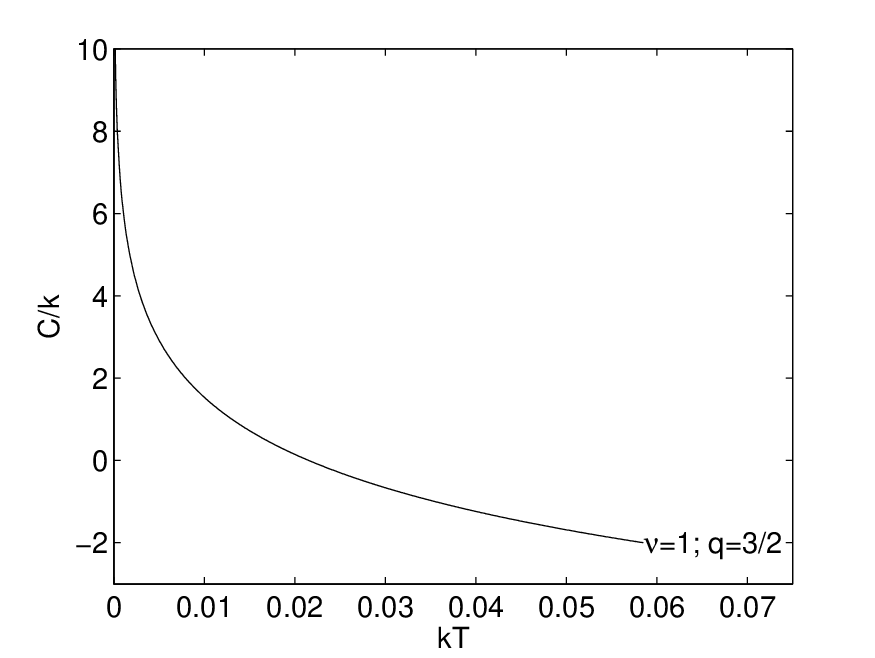}
\vspace{-0.2cm} \caption{One dimension: specific heats at the pole
versus temperature $T$, plotted within the allowed temperature
range.}\label{fig1}
\end{center}
\end{figure}

\newpage
\begin{figure}[h]
\begin{center}
\includegraphics[scale=0.6,angle=0]{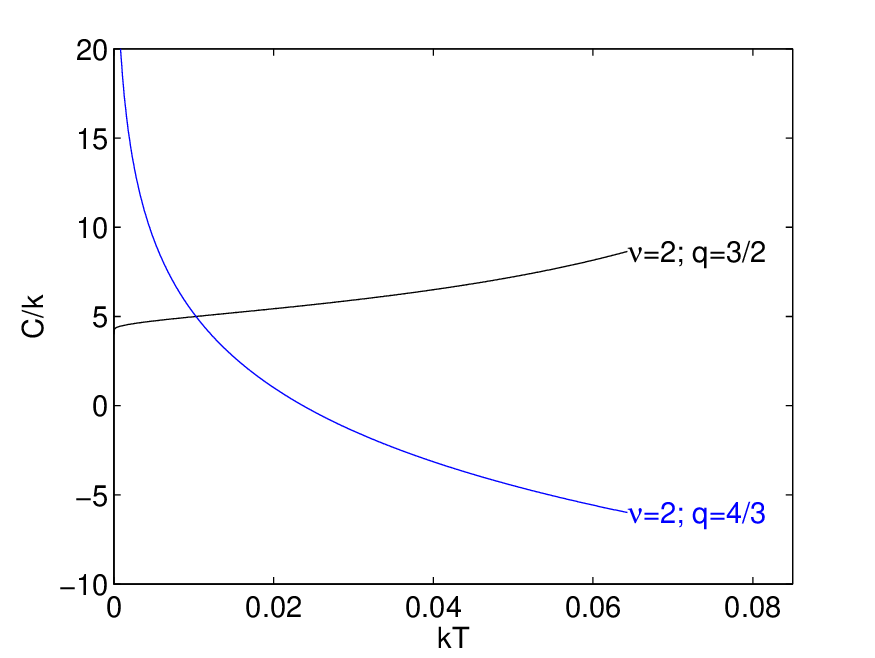}
\vspace{-0.2cm} \caption{Two dimensions: specific heats at the two
poles versus temperature $T$, plotted within the allowed
temperature ranges in the two cases.}\label{fig2}
\end{center}
\end{figure}

\newpage
\begin{figure}[h]
\begin{center}
\includegraphics[scale=0.6,angle=0]{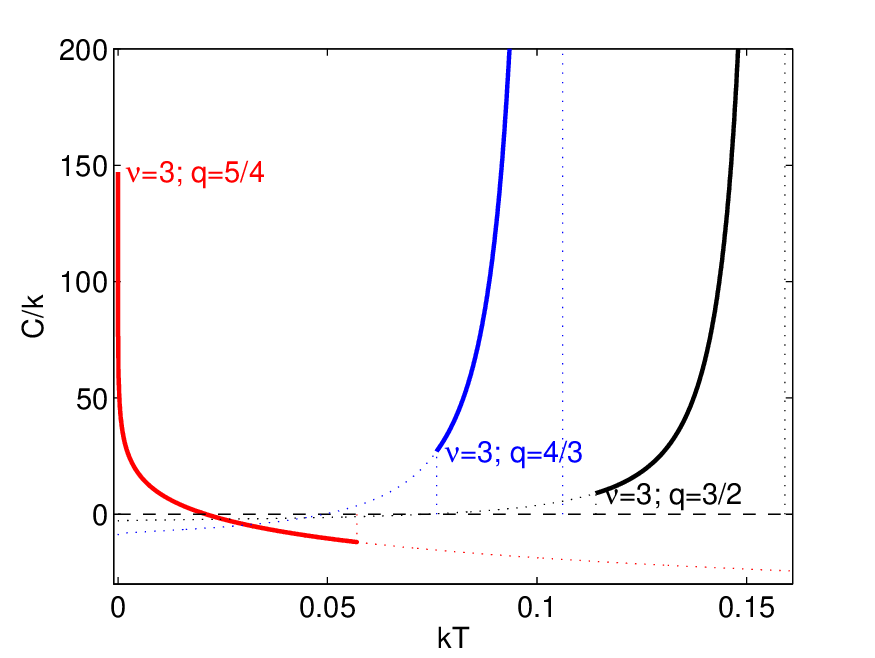}
\vspace{-0.2cm} \caption{Three dimensions: specific heats at the
three poles versus temperature $T$. The vertical lines demarcate
the allowed temperature ranges in the three cases. Dashed lines
are continuations of the $C-$values outside the domains of
validity }\label{fig3}
\end{center}
\end{figure}


\begin{thebibliography}{99}
\bibitem{xxx} The procedure we have employed here  is usually called
``dimensional regularization".
\end{thebibliography}
\end{document}